\documentclass[useAMS,usenatbib,usegraphicx]{mn2e}
\usepackage{amsmath}
\usepackage{amssymb}
\usepackage{graphicx}

\def\simlt{\lower.5ex\hbox{$\; \buildrel < \over \sim \;$}}
\def\simgt{\lower.5ex\hbox{$\; \buildrel > \over \sim \;$}}

\def\beq{\begin{equation}}
\def\eeq{\end{equation}}

\def\bc{_\mathrm{bc}}
\def\cool{_\mathrm{cool}}
\def\coolz{_\mathrm{cool,0}}

\def\gas{_\mathrm{gas}}

\newcommand{\bd}{\begin{displaymath}}
\newcommand{\ed}{\end{displaymath}}
\newcommand{\beqa}{\begin{eqnarray}}
\newcommand{\eeqa}{\end{eqnarray}}

\newcommand{\vbc}{v_{\rm bc}}
\renewcommand{\vec}[1]{{\bf #1}}

\title[Impact of Relative Motion on the First Stars]{Impact of the
Relative Motion between the Dark Matter and Baryons on the First
Stars}

\author[Fialkov, Barkana, Tseliakhovich \& Hirata]
{Anastasia Fialkov$^{1}$\thanks{E-mail: anastasia.fialkov@gmail.com},
Rennan Barkana$^{1}$, Dmitriy Tseliakhovich$^{2}$,
\newauthor Christopher M. Hirata$^3$ \\ $^{1}$ Raymond and
Beverly Sackler School of Physics and Astronomy, Tel Aviv University,
Tel Aviv 69978, Israel \\ $^{2}$ California Institute of Technology,
M/C 249-17, Pasadena, California 91125, USA \\ $^{3}$ California
Institute of Technology, M/C 350-17, Pasadena, California 91125, USA }

\begin{document}
\maketitle

\label{firstpage}

\begin{abstract}
Recently the initial supersonic relative velocity between the dark
matter and baryons was shown to have an important effect on galaxy
formation at high redshift. We study the impact of this relative
motion on the distribution of the star-forming halos and on the
formation redshift of the very first star. We include a new aspect of
the relative velocity effect found in recent simulations by fitting
their results to obtain the spatially-varying minimum halo mass needed
for molecular cooling. Thus, the relative velocities have three
separate effects: suppression of the halo abundance, suppression of
the gas content within each halo, and boosting of the minimum cooling
mass. We show that the two suppressions (of gas content and of halo
abundance) are the primary effects on the small minihalos that cannot
form stars, while the cooling mass boost combines with the abundance
suppression to produce order unity fluctuations in stellar density. We
quantify the large-scale inhomogeneity of galaxies, finding that
$68\%$ of the star formation (averaged on a 3~Mpc scale) is confined
to $35\%$ of the volume at $z=20$ (and just $18\%$ at $z=40$). In
addition, we estimate the redshift of the first star to be $z\sim 65$,
which includes a delay of $\Delta z \sim 5$ due to the relative
velocity.
\end{abstract}


\section{Introduction}

In the present era of ``precision cosmology'' and rapidly advancing
observational capabilities it is important to make precise theoretical
predictions for future observations. Among the major goals of
observational cosmology in the near future are to collect data on
structure at high redshifts (including the first galaxies), detect the
21-cm line of intergalactic hydrogen, and study the cosmic
reionization history. A deep understanding of structure formation on
small scales and at high redshifts is crucial for making reliable
predictions that will help us explore this observational frontier.

The linear perturbation theory of structure formation in the framework
of the flat $\Lambda$CDM model is well understood. It allows us to
follow the evolution of structure starting from tiny
perturbations. The large-scale perturbations are $\mathcal O
(10^{-5})$ of the background quantities at cosmic recombination,
$z\sim 1100$ \citep{WMAP7}, and may have been produced during an early
period of inflation. Structure on the smaller scales on which halos
form evolves nonlinearly. In order to make reliable predictions, it is
important to verify when we can trust the results of the linear
perturbation theory and on which scales the nonlinear effects must be
accounted for.

Linear theory separates different scales, so that each density
perturbation mode at a given wavenumber $k$ evolves
independently. Thus, nonlinear terms that couple the large-scale
velocity to the small-scale density perturbations are neglected in
linear perturbation theory. However, recently it was shown
\citep{Tseliakhovich:2010} that such terms might be of the
same order of magnitude as the linear terms exactly at the time and on
the scales on which the first baryonic objects formed. Specifically,
the photon-baryon coupling before recombination left the dark matter
and baryonic fluids with large relative velocities. These velocities
impede the gravitational perturbation growth on small scales, leading
to a spatially-variable suppression in the abundance of halos
\citep{Tseliakhovich:2010}. Moreover, halos that
later form cannot accrete the gas as it shoots past the collapsing
dark matter \citep{Dalal:2010,Tseliakhovich:2010b}.

In this paper we study the impact of the relative velocities on the
distribution of the star-forming halos at high redshift and on the
redshift of formation of the very first star. In particular, we
include an aspect of the relative velocity effect that has not been
previously accounted for, and which is critical for understanding the
overall impact of the velocities on the distribution of star
formation. Recent small-scale numerical simulations
\citep{Stacy:2011,Greif:2011} found that the relative velocity
substantially increases the minimum halo mass in which stars can form
from gas that cools via molecular hydrogen cooling (The effect of the
velocities has also been simulated by \citet{Maio:2011} and
\citet{Naoz:2011}).

This paper is organized as follows.  In Section~2 we briefly review
the results of \citet{Tseliakhovich:2010} and
\citet{Tseliakhovich:2010b}. In Section~3 we summarize the results of
recent simulations that include the nonlinear effect of the relative
velocity on the formation of the first stars via molecular cooling. We
use the simulation results to find the behavior of the minimal cooling
mass versus redshift and magnitude of the relative velocity. In
Section~4 we study in detail the probability distribution of the gas
fraction in halos at high redshift, separating out and comparing the
importance of the various effects of the bulk velocity. In Section~5
we then estimate the redshift of the very first star accounting for
the relative velocity effect. Finally, in Section~6 we summarize our
results and also give a complete list of differences compared to three
previous papers: \citet{Tseliakhovich:2010}, \citet{Dalal:2010}, and
\citet{Tseliakhovich:2010b}.

Our calculations are carried out in a flat $\Lambda$CDM universe with
cosmological parameters taken from the 7-year WMAP results
(WMAP7+BAO+$H_0$ maximum likelihood fit from \citet{WMAP7}): the dark
matter density today $\Omega_{c,0} = 0.2265$, the baryon density
$\Omega_{b,0} = 0.0455$, the vacuum energy density $\Omega_\Lambda =
0.728$, the Hubble constant $H_0 =70.4$ km s$^{-1}$ Mpc$^{-1}$, and
the spectral index $n_s = 0.967$. We normalize the power spectrum to
give a present value of $\sigma_8 = 0.81$ \citep{WMAP7}. We use the
CAMB-sources linear perturbation code \citep{CAMBsources} to generate
initial conditions at recombination (specifically, at $z = 1020$ and
$z = 970$ in order to obtain the needed derivatives).

\section{Review of the Relative Velocity Effect}

In this section we briefly review the non-linear effect of the
relative velocities between the baryons and dark matter, as discussed
in \citet{Tseliakhovich:2010} and \citet{Tseliakhovich:2010b}, the
latter of which we closely follow in our subsequent calculations.

The initial conditions at recombination include significant relative
velocities between the baryons and the cold dark matter (which we
denote $v\bc$). Before the baryons kinematically decouple from the
radiation (around $z=1100$), they are carried along with the photons,
while the dark matter moves according to the gravitational growth of
fluctuations which has been advancing since matter-radiation equality
($z \sim 3200$). At decoupling, the baryonic speed of sound drops
precipitously, and the relative velocity then becomes a substantial
effect.

In the standard picture of Gaussian initial conditions (e.g., from a
period of inflation), the density and the components of relative
velocity are Gaussian random variables. The velocity and density are
spatially correlated (at different points) since the continuity
equation relates the velocity divergence to the density. Indeed, this
equation gives an extra factor of $1/k$ in the velocity (where $k$ is
the wavenumber), making the velocity field coherent on larger scales
than the density. Specifically, velocity fluctuations have significant
power over the range $k \sim 0.01-0.5$ Mpc$^{-1}$.

The relative velocity is thus coherent on scales smaller than $\sim 3$
comoving Mpc. We therefore analyze probability distributions in such
coherent patches, and refer to the uniform relative velocity within
each patch as the ``bulk'' or ``streaming'' velocity. The magnitude of
the bulk velocity in each coherence patch at recombination is
distributed according to a Maxwell-Boltzmann distribution function:
\beq \label{MBdist}p_{v\bc}(v\bc) = \left(\frac{3}
{2\pi \sigma^2_{ v\bc}}\right)^{3/2}4\pi  v\bc^2
\exp\left(-\frac{3 v\bc^2}{2\sigma^2_{ v\bc}}\right)\ ,\eeq
where $\sigma_{ v\bc}\sim 30 $ km sec$^{-1}$ is the root-mean-square
velocity at recombination. Just like any peculiar velocity, the bulk
velocity $v\bc$ decays as $(1+z)$ with the expansion of the universe.
In addition to the bulk velocity, within each patch there are
small-scale peculiar velocities of the baryons and dark matter related
to the evolution of perturbations (and formation of halos) within the
patch.

As was shown in the above references, inside each coherent region the
linear evolution equations for density and velocity perturbations are
modified. For example, on small scales the nonlinear term in the
continuity equation that couples the local density to the velocity
field, $a^{-1} \bf{v} \cdot \bf{\nabla} \delta$, is comparable to
linear terms such as the velocity term $a^{-1}\bf{\nabla} \cdot
\bf{v}$. The leading contribution of the nonlinear term comes from
the bulk motion ($a^{-1} \bf{v}\bc \cdot \bf{\nabla} \delta$) and this
contribution is then linear in terms of the perturbations within the
patch. As a result, the evolution equations for the perturbations
inside a coherent patch are still linear but dependent on the bulk
$v\bc$. The resulting velocity-dependent terms were previously
neglected but must be included when structure on small scales and at
high redshifts is considered.

The relative velocity effect is particularly important for the
formation of the first stars and galaxies. As the first baryonic
objects try to form, they must do so in a moving background of the
dark matter potential wells. This relative motion means that the dark
matter's gravity must work harder in order to trap the baryons. As a
result, the formation of the first bounded baryonic objects is
delayed. The effect, though, is less relevant for structure formation
today, since the relative velocity decays with time while the typical
mass of galactic host halos increases. However, the relative motion
may shift slightly the positions of the BAO peaks and produce a unique
signature in the bispectrum of galaxies
\citep{Yoo:2011}.

\section{Calibration of the minimum halo cooling mass with
simulations}
\label{CM:FS}

The formation of the first baryonic objects (in particular the first
stars) was an important milestone in the history of the Universe. It
marked the transition between the cold, neutral, metal-free universe
(the epoch called the ``dark '' cosmological ages that started right
after recombination) and the modern ionized, hot, and metal-rich
universe. The formation of the very first stars is expected to be
relatively simple; this is due to the primordial chemistry before
stars produced heavy elements, and the simplified gas dynamics in the
absence of dynamically-relevant magnetic fields and feedback from
luminous objects \citep{Tegmark:1997,Barkana:2001}.

Since molecular hydrogen line emission is the lowest-temperature
coolant in metal-free gas, the first stars are expected to have formed
in halos with total mass above $\sim 10^5$ M$_\odot$
\citep{Tegmark:1997}. More generally, if the mass of a dark matter
halo is higher than a threshold referred to as the minimum cooling
mass ($M\cool$), the collapsing gas is heated to a high enough
temperature that it emits radiation. It then cools and condenses,
allowing a star to form. The threshold can also be described as a
minimum circular velocity ($V\cool$) via the standard relation $V_c =
\sqrt{GM/R}$ for a halo of mass $M$ and virial radius $R$.

This scenario of the earliest star formation has been confirmed by
numerical simulations using both Adaptive Mesh Refinement (AMR) and
Smooth Particle Hydrodynamics (SPH) codes \citep[e.g.,][]{Fuller:2000,
Abel:2001, Bromm:2002, Yoshida:2003, Reed:2005, Yoshida:2006,
Turk:2011}. All these simulations did not account for the initial
relative velocities between the baryons and the dark matter.  We now
summarize two recent SPH simulations \citep{Stacy:2011, Greif:2011}
that studied the impact of the relative streaming velocity $v\bc$ on
the formation of the first stars.

Numerical simulations face a great difficulty at high redshift, since
they must resolve the then-typical tiny galaxies while at the same
time capture the global galaxy distribution which is characterized by
strong fluctuations on surprisingly large scales
\citep{Barkana:2004}. The relative velocities are correlated up to
scales above 100~Mpc, and they are important at high redshifts where
star formation is dominated by very small halos. Cosmological
simulations that cover this range of scales are not currently
feasible.

However, numerical simulations are the best tool for studying the
complex, non-linear formation of halos on small scales. The scales
relevant to the formation of the small halos that host the first
stars are well below the coherence scale of the relative velocity
field. Therefore it is possible to simulate halo formation in small
patches of uniform $v\bc$. The simulations yield the mass reached by a
halo when it first allows a star to form, i.e., when it first contains
a cooling, rapidly-collapsing gas core. The results show a
substantially increased halo mass in regions with a significant
relative velocity. This is a different effect from the suppression of
the amount of gas, which implies a smaller number of stars in the halo
at a given time; instead in this case there is a substantial delay in the
formation of the first star within the halo. Moreover, this effect is
not simply related to the total amount of accreted gas, since in the
cases with a bulk velocity, even if we wait for the halo to accrete
the same total gas mass as its no-velocity counterpart, it still does
not form a star (even within the now deeper potential of a more
massive host halo); the delay is substantially longer than would be
expected based on a fixed total mass of accreted gas. Instead, it
appears that the explanation lies with the internal density and
temperature profiles of the gas, which are strongly affected by the
presence of the streaming motion. A plausible explanation for the
resulting delay in star formation is that the first star forms from
the gas that would have accreted early and formed the dense central
cores in which stars form; this gas tries to accrete early (when
$v\bc$ is still very large) into a still-small halo progenitor, so it
is affected most strongly by the suppression of gas accretion due to
the bulk velocity.

The simulations yield a minimum halo cooling mass at various
redshifts, so we fit the results to find the dependence of the minimum
halo mass on the redshift of collapse and on the bulk velocity,
$v\bc$, in the patch. This will then allow us to study the effect of
the relative velocity on the formation of the first stars using
statistical methods that average over large cosmological regions that
cannot be directly simulated.

\citet{Stacy:2011} and \citet{Greif:2011} state
apparently contradictory conclusions, one claiming a negligible effect
on star-forming halos and the other a large effect. In order to
meaningfully compare their results, it is important to put them both
on the same scale. We express the cooling threshold as a halo circular
velocity, since simulations (cited above) without the bulk velocity
find an approximately redshift-independent threshold $V\coolz $; this
is naturally expected since molecular cooling turns on essentially at
a fixed gas temperature, and the halo circular velocity determines the
virial temperature to which the gas is heated. Thus, the limit of zero
bulk velocity simply gives a fixed threshold $V\coolz $. When we add
the relative velocities, in principle the minimum circular velocity in
a patch may be a separate function of two parameters, the redshift $z$
and the bulk velocity at halo formation $v\bc(z)$. The history of
$v\bc$ at earlier redshifts cannot introduce additional parameters,
since given both $z$ and $v\bc(z)$, the full history of $v\bc$ is
determined, i.e., at any other redshift $z'$, $v\bc(z')=v\bc(z) \times
(1+z')/(1+z)$.

Consider now the limit of a very high bulk velocity, $v\bc(z) \gg
V\coolz $, so that the effect of $V\coolz $ is negligible. For
simplicity, consider for a moment a constant $v\bc$ versus redshift,
fixed at its final value $v\bc(z)$ at the halo formation redshift $z$.
In this case there is only one velocity scale in the problem. As in a
Jeans mass analysis, in the reference frame of a collapsing dark
matter halo with a circular velocity $V_c$, clearly gravity will be
able to pull in the gas (which streams by at the velocity $v\bc(z)$)
if $V_c \ga v\bc(z)$. Now, in the real case where $v\bc(z')$ is higher
during the formation of the halo, we would expect to get a threshold
that is higher than $v\bc(z)$, but by a fixed factor, because the
physics is scale-free: on one side, $v\bc$ scales in a simple way with
redshift, and on the other side, halo formation (in the high-redshift,
Einstein de-Sitter universe) also scales in a simple way, as we know
from spherical collapse; e.g., turnaround for a halo that forms at
redshift $z$ always occurs at $z'$ where $1+z' = 1.59 (1+z)$ so that
$v\bc(z') = 1.59 v\bc(z)$. The only new scale that enters is from
$v\bc$ at recombination, but as long as we consider halos that
form long after recombination, this should be insignificant.

Thus, the threshold circular velocity $V\cool$ should change
continuously between two limits, $V\cool=V\coolz $ when $v\bc(z) \ll
V\coolz $, and $V\cool=\alpha v\bc(z)$ when $v\bc(z) \gg V\coolz $ (in
terms of a fixed, dimensionless parameter $\alpha$). When $V\cool$ is
expressed as a function of $v\bc(z)$, there is no additional
dependence on $z$ in these two limits, so we might naturally expect
this to be true in the intermediate region as well. Indeed, the above
argument suggests more generally that halo formation and $v\bc(z)$
scale together so that the effect of the bulk velocity should not
depend separately on redshift; also the effect of molecular cooling is
a redshift-independent threshold. Thus, when both effects act
together, the result should still depend on just one parameter.

We expect the dependence on velocity to be smooth and well-behaved for
vector $\vec{v}\bc(z)$ near zero, i.e., as a function of the velocity
components. This suggests a quadratic dependence on $\left[ v\bc(z)
\right]^2 = \left[ \vec{v}\bc(z) \right]^2$ rather than, e.g., a
linear dependence on $v\bc(z)$. We thus propose a simple ansatz for
the minimum cooling threshold of halos that form at redshift $z$:
\begin{equation} V\cool(z) =\left\{ V\coolz ^2+\left[\alpha v\bc(z)
\right]^2 \right\}^{1/2}\ .
\end{equation}
The dependence of the circular velocity $V\cool$ on redshift only
through the final value $v\bc(z)$ implies that the star-formation
threshold in a patch with a statistically rare, high value of $v\bc$
at low redshift is the same as the threshold in a patch with the same
(but now statistically more typical) value of $v\bc$ at high
redshift. This should be the case during the era of primordial star
formation, before metal enrichment and other feedbacks complicate
matters.

\begin{figure}
\includegraphics[width=3.4in]{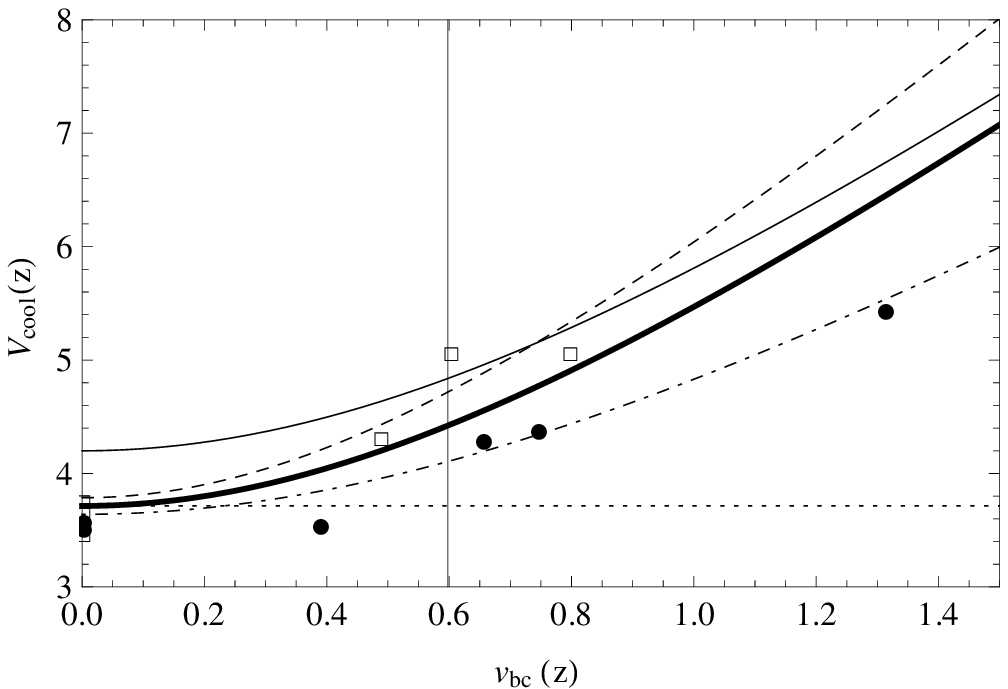}
\includegraphics[width=3.4in]{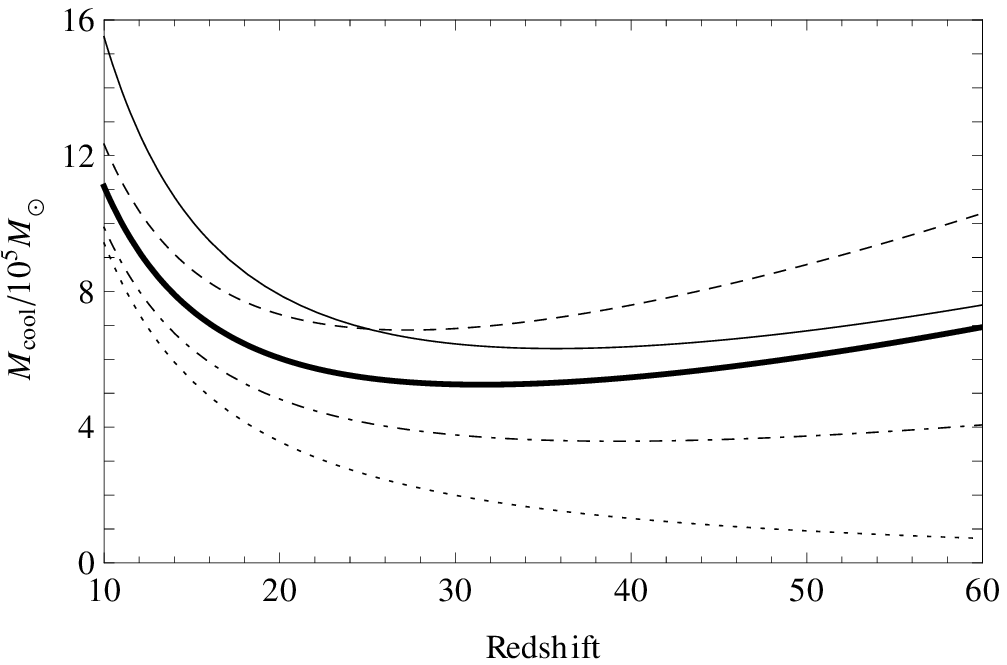}
\caption{\label{Fig:Vc}  {\bf Top panel}: The minimum halo circular
velocity for gas cooling via molecular hydrogen versus the bulk
velocity $v\bc(z)$ when the halo virializes. Data are taken from
\citet{Stacy:2011} ($\bullet$) and \citet{Greif:2011} ($ \square $).
We show our fits to each set of simulation results (dot-dashed and
dashed, respectively). We also show our ``optimal'' fit to SPH
simulations (thick solid line), the ``fit'' to AMR simulations
(regular solid line), and the case of no streaming velocity (dotted
line, based on our optimal fit). The vertical solid line marks the
root-mean-square value of $v\bc(z)$ at $z = 20$. {\bf Bottom panel}:
We show the minimum halo mass for molecular cooling versus redshift,
in a patch with the root-mean-square value of $v\bc(z)$ at each
redshift $z$, for each of the fits from the top panel; in particular,
we show (dotted line) the case of no relative motion based on our
optimal fit (i.e., $V\cool = V\coolz = 3.714$ km sec$^{-1}$).}
\end{figure}

We summarize the results of the two simulations together with the best
fits to each of them (with $V\coolz $ and $\alpha$ as free parameters)
in Figure~\ref{Fig:Vc} (top panel). We obtain four data points from
\citet{Stacy:2011} with non-zero velocities (and two more at
$v\bc(z)=0$), and three points from \citet{Greif:2011} (plus three
more at $v\bc(z)=0$). The best-fit parameters are: (1) $ V\coolz  =
3.640$ km sec$^{-1}$ and $\alpha = 3.176$ for the results of
\citet{Stacy:2011}; (2) $ V\coolz  = 3.786$ km sec$^{-1}$ and
$\alpha = 4.707$ for \citet{Greif:2011}.

We note that despite the small numbers of halos, we would not
necessarily expect as large a scatter in the measured $V\cool(z)$ as
in other measurements of halo properties; for example, in a sample
with a large number of halos of various masses at each redshift, we
would expect a large range of redshifts for the first star formation
within a halo, but if we only take halos that first formed a star at a
given redshift $z$, their masses at $z$ might span a narrow range, all
near the minimum cooling mass for that redshift (since any halo well
above the cooling mass at $z$ would already have formed a star
earlier). In any case, our ansatz fits each set of simulation results
reasonably well, but there is some scatter and also a systematic
difference between the two sets (with \citet{Greif:2011} indicating a
stronger effect of the bulk velocity). Due to the small number of
simulated halos, it is difficult to separate the possible effects of
different numerical resolutions, other differences in the
gravitational or hydrodynamical solvers, and real cosmic scatter among
halos. Given the systematic offset, we do not simultaneously fit both
sets of points, but instead average the best-fit parameters of the two
SPH simulation sets. We mostly use this fit, which we refer to as our
{\it optimal fit}, in the following sections:
\begin{equation} V\cool(z) =\left\{ \left( 3.714\ {\rm km/s}
\right)^2+\left[4.015 \cdot v\bc(z) \right]^2\right\}^{1/2}\ .
\label{eq:opt}
\end{equation}

There is some discrepancy in the value of $V\coolz $ found in AMR and
SPH simulations. In order to test the full current uncertainty range
including different types of simulations, we also consider the average
value $ V\coolz  \sim 4.2$ km sec$^{-1}$ found in AMR simulations
\citep{Yoshida:2006,Turk:2011}. Thus, we combine this value of
$ V\coolz  $ with $\alpha$ from our optimal fit to obtain what we refer
to as a ``fit'' to AMR simulations. In other words, we assume that the
discrepancy between the two simulation methods is only in the cooling
process (due to systematic entropy differences in dense cores), but
that they would agree on the effect of the bulk motion. Regardless of
which fit we use, Figure~\ref{Fig:Vc} shows that the relative motion
has a large effect on the minimum circular velocity.

The implications for the minimum cooling mass as a function of
redshift are also shown in Figure~\ref{Fig:Vc} (bottom panel). In a
patch with no relative motion, the mass drops rapidly with redshift,
since at higher redshift the gas density is higher and a given halo
mass heats the infalling gas to a higher virial temperature. However,
in a region at the root-mean-square value of $v\bc$ \footnote{Since
$v\bc$ decays as $1+z$ throughout the universe, a patch that has the
root-mean-square value of $v\bc$ at one redshift will have the
root-mean-square value of the relative velocity at every redshift, and
in particular $v\bc =30$ km sec$^{-1}$ at recombination.}  the higher
bulk velocity at high redshift implies that a higher halo mass is
needed for efficient molecular cooling. In particular, at redshift 20
a patch with $v\bc = 0$ will form stars in $3.6 \times 10^5$ M$_\odot$
halos, while a patch with the root-mean-square value of $v\bc$ has a
minimum cooling mass of $6.0 \times 10^5$ M$_\odot$ according to the
optimal fit, or a range of $(4.8 - 7.3) \times 10^5$ M$_\odot$ from
the other fits. At $z=60$ these numbers become $7.2 \times 10^4$
M$_\odot$, $7.0 \times 10^5$ M$_\odot$, and $(4.1 - 10.3) \times 10^5$
M$_\odot$, respectively. In patches with low bulk velocity we expect
stars to form earlier, since the halos with lower masses are more
abundant and form earlier in the hierarchical picture of structure
formation. This is the basis of the discussion that follows.

\section{Gas Fraction in the First Bound Baryonic Objects}

\subsection{Global average}

The population of gas-filled halos at high redshift divides naturally
into two major categories. The first category consists of large halos
in which the gas can cool (via molecular hydrogen cooling); these are
presumed to be the sites of formation of the first stars, and are
obviously most important since the stellar radiation is in principle
observable, and it also produces feedback on the intergalactic medium
and on other nearby sites of star formation. Also interesting, though,
is the second category, namely the smaller halos (``minihalos'') in
which the gas accumulates to roughly virial density and yet cannot
cool. The latter may affect the epoch of reionization by acting as a
sink for ionizing photons \citep[e.g.,][]{Haiman01, BL02, Iliev05,
Ciardi05} and may generate a 21-cm signal from collisional excitation
of H{\sc\,i} \citep[e.g.,][]{Iliev03, FO06}.

In this subsection we apply the result we found for the minimum
cooling mass to find the redshift evolution of the gas fraction in
these two categories. In the following subsections we explore the
probability distribution function (PDF) of the gas fraction, beginning
with its dependence on the bulk velocity. In addition, though, in each
patch of coherent velocity the mean density is slightly different,
varying as a result of random density fluctuations on scales larger
than the patch size. We thus also study the full PDF as determined by
the joint dependence of the gas fraction in halos on the bulk velocity
and the local overdensity in each patch.

Following \citet{Tseliakhovich:2010b} we find the fraction of the
baryon density contained in halos with mass larger than the minimum
cooling mass $M\cool$
\beq f\gas(>M\cool)=\int_{M\cool}^{\infty} \frac{M}{\bar\rho_0}
\frac{dn}{dM}\frac{f_g(M)}{f_b}dM\ ,\eeq
where $ \bar\rho_0$ is the mean matter density today, $dn/dM$ is the
comoving abundance of halos of mass $M$, $f_b\equiv
\Omega_b/\Omega_m$ is the mean cosmic baryon fraction and $f_g(M)$ is
the fraction of the total halo mass which is in the form of gas. The
gas fractions $f_g(M)$ depend on the filtering mass, which measures
the scale at which the baryon fluctuations differ substantially from
those in the dark matter. In each patch, the filtering mass depends on
the bulk velocity, and thus so do the gas fractions. Since the baryons
contribute to the total power spectrum, the halo abundance $dn/dM$
(which depends on fluctuations in the total matter density) varies as
well with $v\bc$. We use the halo mass function of
\citet{Shetht:1999}. We refer the reader to
\citet{Tseliakhovich:2010b} for the full details.

\begin{figure}
\includegraphics[width=3.4in]{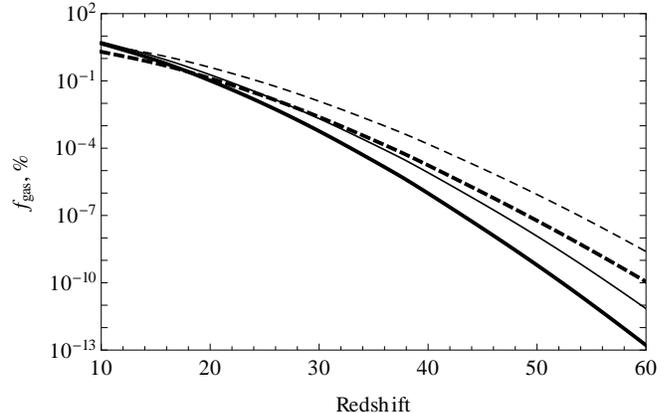}
\caption{\label{Fig:Vc2} The global mean gas fraction in
star-forming halos (solid curves) and in minihalos, i.e., halos below
the cooling threshold (dashed curves). The results, based on our
optimal fit (eq.\ \ref{eq:opt}) are shown after averaging over the
distribution of relative velocity (thick curves), or in the case of no
relative motion, i.e., for $v\bc(z)=0$ (thin curves).}
\end{figure}

We begin by recalculating some of the results of
\citet{Tseliakhovich:2010b}. We show in Figure~\ref{Fig:Vc2} the
redshift evolution of the globally averaged gas fraction in
star-forming halos or in gas minihalos. Compared with Figure~8 of
\citet{Tseliakhovich:2010b}, our gas fractions are substantially lower,
e.g., the gas fraction in halos above the minimum cooling mass is
lower by a factor of $\sim 3$ at redshift $z = 20$. This is due to our
higher $M\cool$ and lower power spectrum normalization (see Section~6
for a full discussion of our differences with previous papers). Note
that the gas fraction in halos above the minimum cooling mass is
proportional to the stellar mass density, assuming a fixed star
formation efficiency (averaged over each 3~Mpc patch).

In general, the importance of the relative velocities increases with
redshift. Comparing the two categories of halos, we find that the
relative suppression of the minihalos is larger than that of the
star-forming halos at low redshift; however, the relative suppression
of the star-forming halos increases faster with redshift, and
eventually it becomes larger than that of the minihalos (beyond $z
\sim 50$). At $z=20$, the bulk velocities reduce the mean gas
fraction in star-forming halos by a factor of 1.8 and that
in minihalos by 3.1.

\begin{figure}
\includegraphics[width=3.4in]{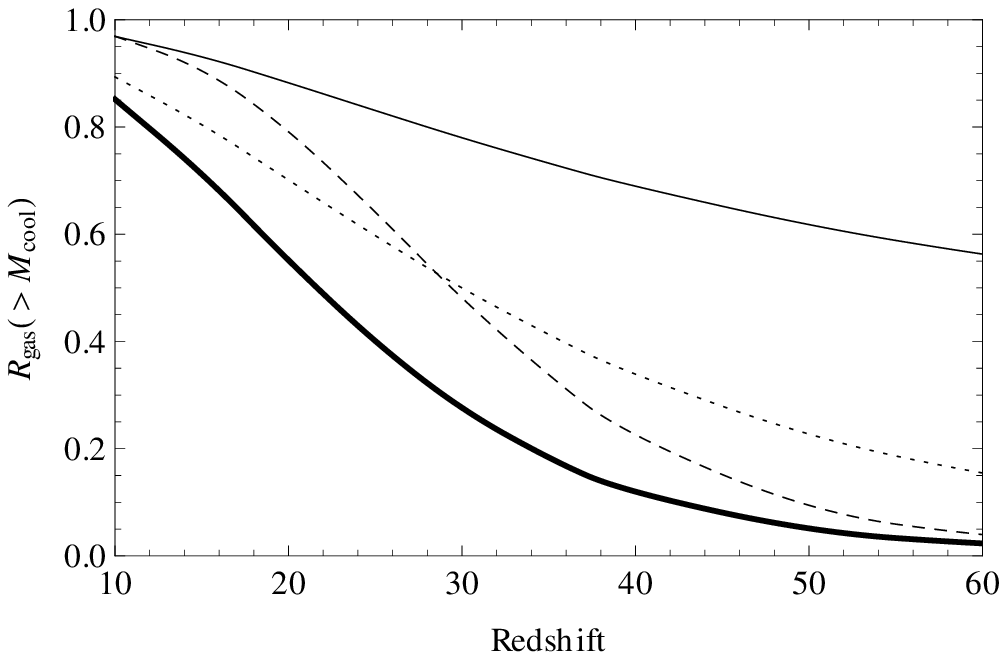}
\includegraphics[width=3.4in]{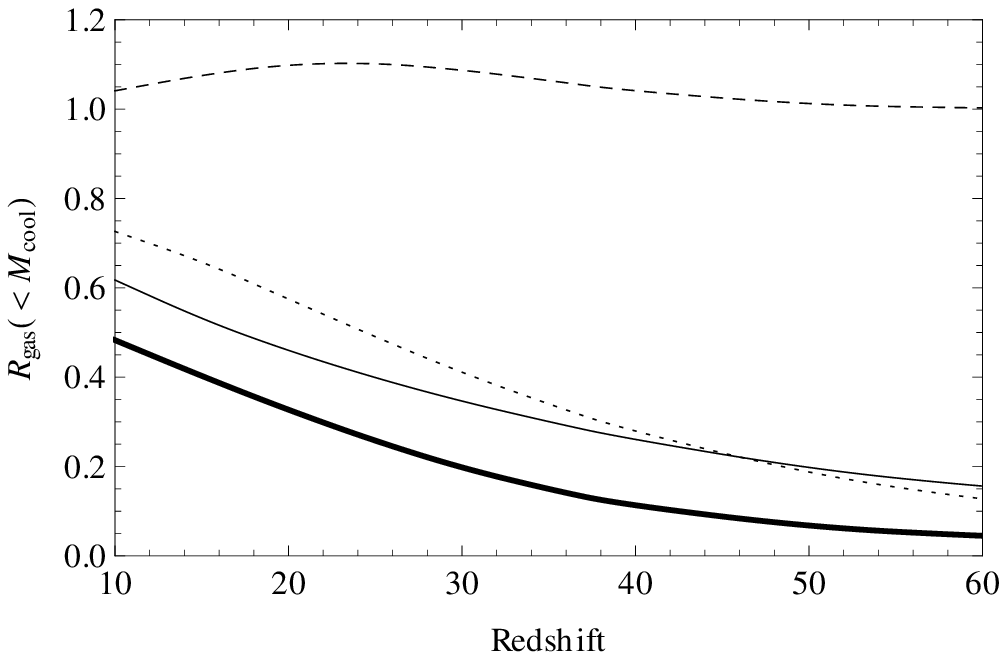}
\caption{\label{Fig:Vc2rat} The ratio (compared to
the $v\bc = 0$ case) by which the bulk velocities change the global
mean gas fraction in halos above the cooling mass (top panel) and in
star-less minihalos (bottom panel). We consider four different cases:
the full effect of the velocities (thick solid curves); the effect of
$v\bc$ in boosting the cooling mass only (dashed curves); the effect
of $v\bc$ in suppressing the halo abundance only (dotted curves); and
the effect of $v\bc$ in suppressing the gas fraction only (thin solid
curves).}
\end{figure}

Unlike previous studies, in our calculations the relative velocities
produce three separate effects: suppression of the halo abundance
($dn/dM$), suppression of the gas content within each halo
($f_g(M)\,$), and boosting of the minimum cooling mass (through
$V\cool(z)\,$). In order to gain a better physical understanding, and
for easier comparison with previous papers, we investigate the
relative importance of each effect in Figure~\ref{Fig:Vc2rat}. For the
star-forming halos, the suppression of gas content is always the least
significant effect (e.g., suppression by a factor of 1.13 on its own
at $z=20$), while the cooling mass boost is most important above
$z=28.5$ (factor of 1.26 on its own at $z=20$), and the halo abundance
cut is most important at lower redshifts (factor of 1.43 on its own at
$z=20$). For the minihalos, the boosting of the minimum cooling mass
acts as a (small) positive effect, since it moves gas from the
star-forming to the minihalo category (e.g., boost by a factor of 1.10
on its own at $z=20$), while the other two effects are larger and
comparable (e.g., at $z=20$ the suppression of gas content would give
a reduction by a factor of 2.17 on its own, and the halo abundance cut
would give a suppression factor of 1.74).

\subsection{Inhomogeneous gas fraction due to the dependence
on the relative velocity}

The gas fractions shown in Figures~2 and 3 are globally
averaged. However, in reality the universe is highly inhomogeneous on
small cosmological scales. We can divide it into patches that have
various bulk velocities and densities. In this section we consider
just the variation with velocity, i.e., averaged over all density
fluctuations. In other words, we look at the contribution of velocity
fluctuations to fluctuations in the gas fraction in halos. If we
consider patches that are still small enough to have a coherent $v\bc$
(e.g., cubes of 3 comoving Mpc on a side), then the absolute value of
the bulk velocity in each one follows a Maxwell-Boltzmann distribution
(eq.~\ref{MBdist}).

\begin{figure}
\includegraphics[width=3.4in]{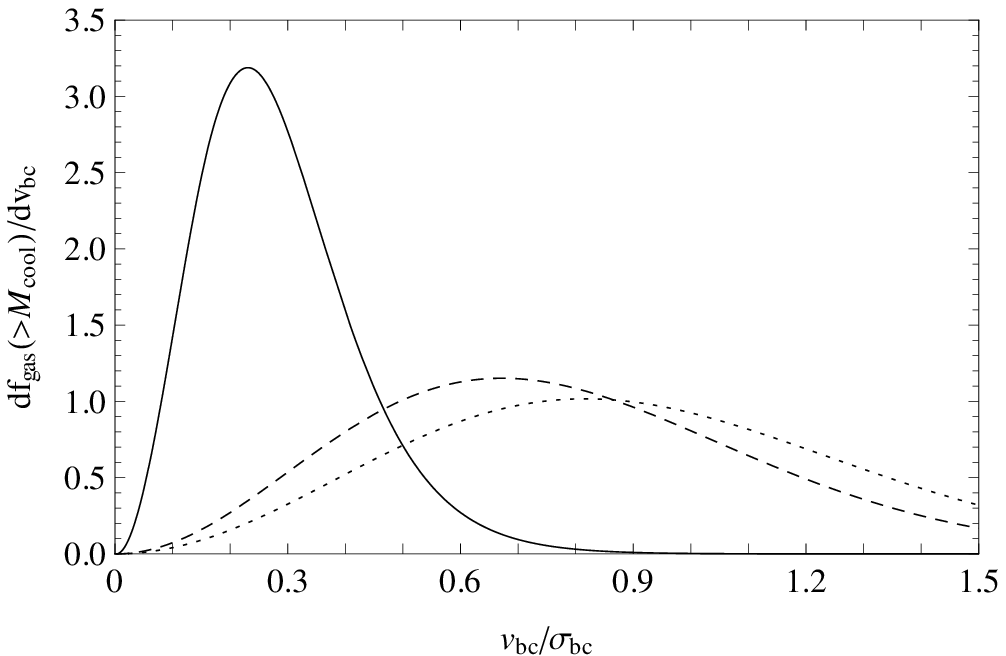}
\includegraphics[width=3.4in]{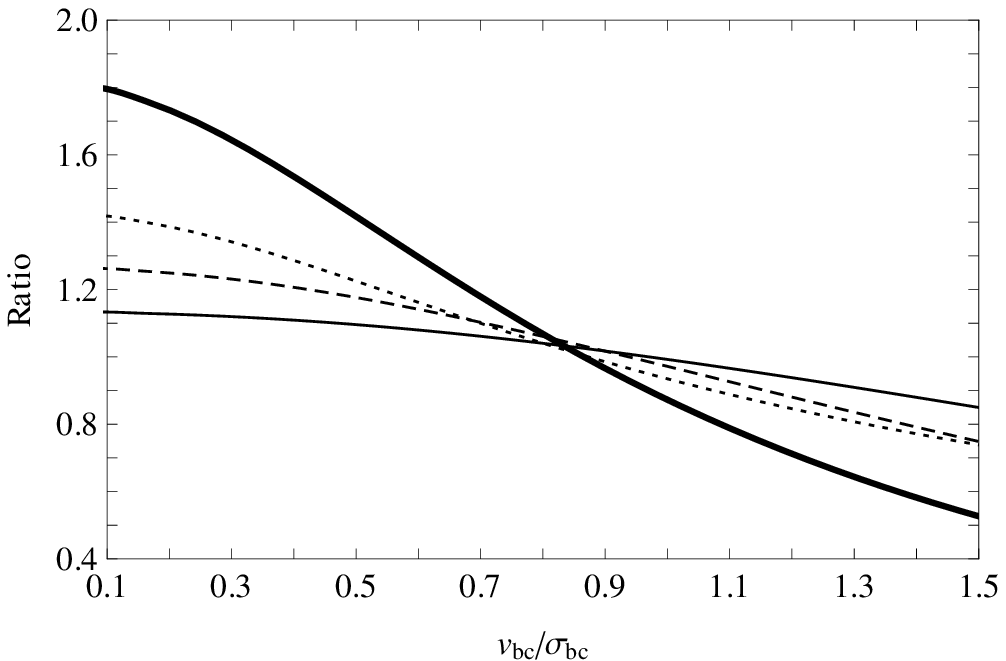}
\caption{\label{Fig:VcPdf}
{\bf Top panel}: The relative contribution of regions with a given
streaming velocity to the global gas fraction in halos above the
cooling mass, i.e., $df_{\rm gas}(>M\cool)/dv\bc$ normalized to an
area of unity. The dependence is shown for $z=60$ (solid curve) and
$z=20$ (dashed curve). We also show the Maxwell-Boltzmann distribution
of the bulk velocity (dotted curve). The velocity is expressed in
units of its root-mean-square value $\sigma_{ v\bc}$. {\bf Bottom
panel}: The ratio at $z=20$ between the quantity shown in the top
panel (the relative contribution of regions with a given streaming
velocity to the gas fraction in star-forming halos) and the
Maxwell-Boltzmann distribution.  If star formation were independent of
bulk velocity, this ratio would equal unity. We consider this ratio
for the same four cases as in Figure~\ref{Fig:Vc2rat}: the full
velocity effect (thick solid curve), the boost in the cooling mass
only (dashed curve). the suppression of halo abundance only (dotted
curve), and the suppression of the gas fraction only (thin solid
curve).}
\end{figure}

Consider the contributions of patches of various velocities to the
total amount of star formation. At a given redshift, the gas fraction
in star-forming halos is lower in the patches with a high value of the
relative velocity, because all three velocity effects (see the
previous subsection) tend to reduce this gas fraction. On the other
hand, patches with zero bulk velocity do not contribute much, simply
because they are rare. As shown in the top panel of
Figure~\ref{Fig:VcPdf}, the most common bulk velocity is $v\bc \sim
0.82 \sigma_{v\bc}$, where $v\bc$ and $\sigma_{v\bc}$ are both
measured at the same redshift (recombination or any other $z$). If the
stellar density were independent of the bulk velocity, then the
contribution of regions of various velocities would be proportional to
the velocity PDF. Instead, the velocity suppression effect shifts the
contribution to stellar density (assumed proportional to the gas
fraction in star-forming halos) towards lower $v\bc$, with the
relative change (compared to the Maxwell-Boltzmann distribution)
increasing strongly with redshift. Thus, the biggest contribution to
stellar density comes from $v\bc = 0.67 \sigma_{v\bc}$ patches at
$z=20$, and from $v\bc = 0.23 \sigma_{v\bc}$ patches at $z=60$. We
compare the contributions of the three separate effects of the
velocity to the shift in the distribution of star formation
(Figure~\ref{Fig:VcPdf}, bottom panel). As in the top panel of
Figure~\ref{Fig:Vc2rat}, we find that the suppression of halo gas
content has the least significant effect on star-forming halos at
$z=20$ (typically, a $\sim 10\%$ effect on the distribution), while
the other two effects (halo abundance suppression and cooling mass
boost) have a $\sim 20-30\%$ effect each.

\begin{figure}
\includegraphics[width=3.4in]{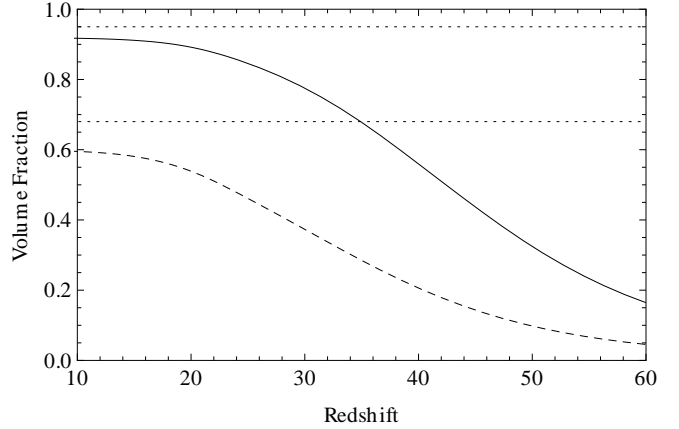}
\caption{\label{Fig:VcPdf2}
The fractional volume of the universe that contains $68\%$ (dashed
curve) or $95\%$ (solid curve) of the star-forming halos as a function
of redshift, where we consider just the contribution of velocity
fluctuations to the inhomogeneity of star formation on 3~Mpc scales.}
\end{figure}

Thus, at the highest redshifts, the star formation is concentrated in
low-velocity regions which are rare, i.e., at the low-probability
$v\bc^2$ end of the Maxwell-Boltzmann distribution function. The
universe at these epochs is very inhomogeneous, with a few bright
regions filled with stars, while in all other regions the relative
velocity is too high to allow significant star formation. As the
universe expands, the relative velocity decays, and in more and more
patches across the universe the relative velocity drops enough to
allow for star formation. As a result, the stellar distribution
becomes increasingly homogeneous. To quantify the degree of
inhomogeneity caused by the dependence of stellar density on the bulk
velocity, we plot the fraction of the volume of the universe (at
lowest velocity, i.e., at highest stellar density) that contains
$68\%$ or $95\%$ of the star-forming halos
(Figure~\ref{Fig:VcPdf2}). The effect of volume concentration is mild
at $z=20$ ($68\%$ of the stars are in $54\%$ of the volume, and $95\%$
in $89\%$ of the volume), while it becomes very strong at $z=60$
($68\%$ of stars in $4.6\%$ of the volume, and $95\%$ in $16\%$ of the
volume).

\subsection{Inhomogeneous gas fraction due to velocity and density
fluctuations}

In order to quantify the full degree of inhomogeneity and
concentration of star formation, we must include the effect of density
fluctuations as well. In this section we thus consider the full PDF of
the halo gas fraction within 3~Mpc patches, where the fluctuations
result from a combination of the relative velocity distribution
considered in the previous section and density
fluctuations. Specifically, the average density in a patch varies due
to fluctuations on scales larger than its size. This average density
follows a Gaussian distribution and is independent of the relative
velocity within the same patch.

\begin{figure}
\includegraphics[width=3.4in]{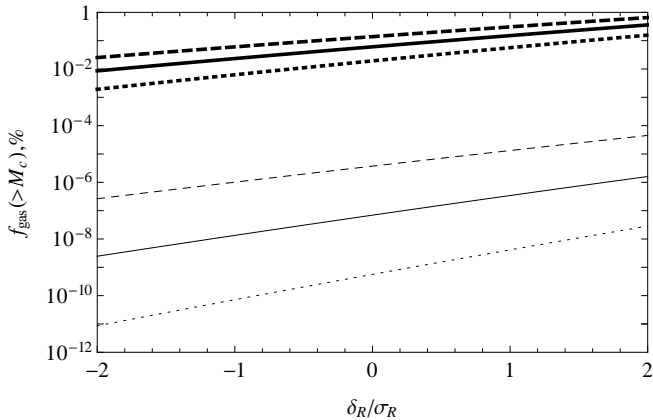}
\caption{\label{Fig:VcDelta}
The percentage of the gas fraction in star-forming halos at redshifts
$z= 20$ (thick curves) and $z=40$ (thin curves) as a function of the
average overdensity $\delta_R$ in the 3~Mpc patch (normalized by its
root-mean-square value $\sigma_R$), for various values of the relative
velocity: no relative motion (dashed), $v\bc = \sigma_{v\bc}$ (solid)
and $v\bc = 2\sigma_{v\bc}$ (dotted).}
\end{figure}

To find the modified halo mass function within a patch of a given
overdensity $\delta_R$ and bulk velocity $v\bc$, we use the hybrid
prescription (which combines the \citet{Shetht:1999} mass function
with the extended Press-Schechter model) introduced by
\citet{Barkana:2004} and generalized by \citet{Tseliakhovich:2010b} to
include $v\bc$.  The dependence of the gas fraction in halos above the
cooling mass on the two independent variables is illustrated in
Figure~\ref{Fig:VcDelta}.  The dependence on both $\delta_R$ and
$v\bc$ (each measured in terms of its root-mean-square value) is
stronger at higher redshifts. At a given redshift, the dependence on
$\delta_R$ is stronger (i.e., the slope is higher) when $v\bc$ is
higher, since in this case the large halos (above the high cooling
mass) are rarer and their abundance is more sensitive to the
overdensity of the patch. If we consider the total range between 0 and
2~$\sigma$, we find that density and velocity fluctuations make
comparable contributions to the star-formation fluctuations on the
3~Mpc scale. The relative importance of velocity increases with
redshift and it will also increase if we consider larger scales. Even
at $z=20$ the velocity causes order unity fluctuations in the stellar
density, and these fluctuations should be present at the large
(100~Mpc) scales spanned by the velocity correlations.

\begin{figure}
\includegraphics[width=3.4in]{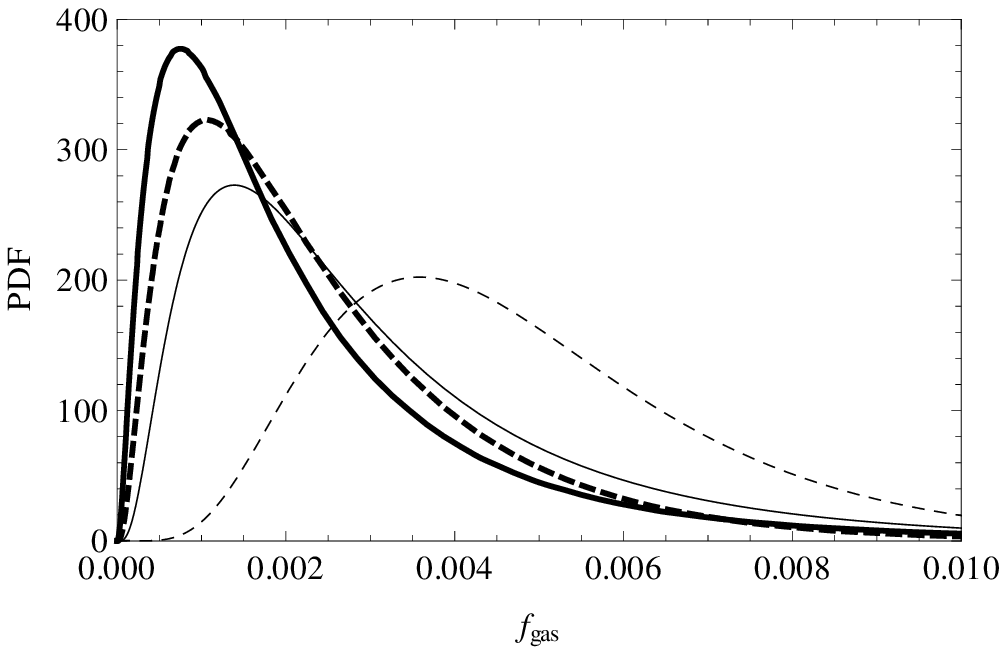}
\includegraphics[width=3.4in]{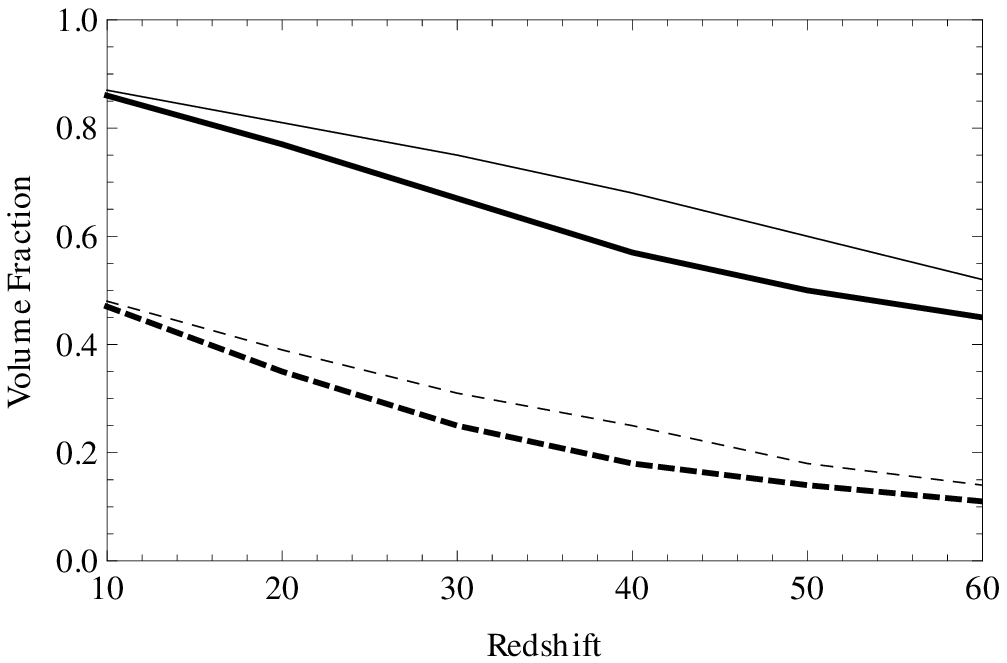}
\caption{\label{Fig:FPdf}
{\bf Top panel}: The full probability distribution function (PDF) of
the gas fraction at redshift $z = 20$. We show the PDF of the gas
fraction in halos above the cooling mass (solid curves) and the PDF of
the gas fraction in star-less minihalos (dashed curves). We consider
two cases: randomly distributed $v\bc$ and $\delta_R$ (thick curves),
and $v\bc = 0$ but random $\delta_R$ (thin curves). {\bf Bottom
panel}: The fractional volume of the universe that contains $68\%$
(dashed curves) and $95\%$ (solid curves) of the star forming halos,
where we consider the full PDF in 3~Mpc patches. In each case we
consider including the relative motion (thick curves) or not ($v\bc =
0$, thin curves).}
\end{figure}

The resulting full PDF of the halo gas fraction is shown in
Figure~\ref{Fig:FPdf} (top panel), both for the star-forming halos,
and the star-less gas minihalos. The main effect of the bulk
velocities is to shift the distributions towards lower gas
fractions. At redshift 20, the effect is larger on the minihalos. In
Figure~\ref{Fig:FPdf} (bottom panel) we show the fraction of the
volume of the universe (at the high gas fraction end of the full PDF)
that contains $68\%$ or $95\%$ of the stars, with and without the
velocity effect.

The volume concentration of star formation is a result of a complex
interplay of the two sources of fluctuations. The global star
formation is highest in the rare regions with both low bulk velocity
and high overdensity, but more generally, one of these can compensate
for the other. The effect of $v\bc$ on star-forming halos vanishes by
$z\sim 10$, in agreement with our previous results, leaving just the
effect of the local density. Even at somewhat higher redshifts (up to
$z \sim 35$), the concentrating effect of the velocities on their own
(Figure~\ref{Fig:VcPdf2}) remains weaker than that of the densities
alone (no-velocity case in Figure~\ref{Fig:FPdf}), so at these
redshifts the full case is dominated by the densities, and the
concentrating effect of density is enhanced by including the
velocities (which steepen the dependence on density:
Figure~\ref{Fig:VcDelta}). At redshifts above $\sim 35$, velocities
dominate, and then including the density fluctuations (compared to
averaging over them at each velocity) actually reduces the
concentration since it allows low-velocity regions to contribute
relatively more volume with high gas fractions (due to the steeper
density dependence at high bulk velocity).

Specifically, at $z=20$, density fluctuations alone (i.e., setting
$v\bc=0$) would concentrate $68\%$ of the stars into $39\%$ of the
volume and $95\%$ into $81\%$ of the volume. The addition of the bulk
velocity provides a mildly increased concentration into $35\%$ and
$77\%$ of the volume, respectively. At redshift 60 the results are
that $68\%$ of the stars are in $11\%$ of the volume and $95\%$ in
$45\%$ (which is higher than in Figure~\ref{Fig:VcPdf2}), compared to
$14\%$ and $52\%$ of the volume, respectively, at zero bulk velocity.
The effect of the velocities should be more clearly apparent on scales
larger than our 3~Mpc pixels, i.e., in addition to the small
additional concentration that they cause (as seen in
Figure~\ref{Fig:FPdf}), their effect is to redistribute the
star-forming regions to produce larger coherent regions of either high
star formation or low star formation (voids).

We note that the assumption that the local overdensity on large scales
$\delta_R$ and the streaming velocity $v\bc$ are statistically
independent is not perfectly accurate. A patch with a high local
overdensity has expanded less than other patches, so that the peculiar
velocity $v\bc$ has not declined as much compared to the expansion.
Indeed, we expect that $\vbc \rightarrow \vbc(1+\delta_R/3)$. However,
we have found that this correction makes only a small difference to
the PDF (up to a $4\%$ relative error at $z = 60$, and less at lower
redshifts).

\section{The First Star}
\label{FS}

In the previous sections we have discussed the conditions needed to
initiate star formation. The main condition is that the halo mass must
be large enough to allow molecular cooling. Given a large enough
initial density fluctuation, a halo with a sufficiently large mass
will form relatively early. The very first stars depend on extremely
rare fluctuations, hence we need to average over the volume of the
observable universe, $(14~\textrm{Gpc})^3$, in order to have the full
statistical range needed to accurately estimate the formation time of
the first star.

Due to computational limitations, numerical simulations can form stars
only in a very limited cosmological context. For instance,
\citet{Greif:2011} studied star formation in a $(500~\textrm{kpc})^3$
volume and \citet{Stacy:2011} were limited to $(100~ \textrm h^{-1}
\textrm{kpc})^3$. In a small volume the chance of getting a rare high
density fluctuation is quite small. Therefore the formation redshift
of the first stars in simulations is greatly underestimated, with most
simulations forming their first star below redshift 30. The highest
redshift where a star has formed in a simulation is $z = 47$
\citep{Reed:2005}.

\citet{Naoz:2006} first applied these statistical considerations
in order to predict the redshift of the first observable star (i.e.,
in our past light cone) analytically. They estimated the redshift of
the first star to be $z = 65$, using the 3-year WMAP set of
cosmological parameters
\citep{Spergel:2006} and assuming a minimum circular velocity for
cooling of $V\cool = 4.5$ km sec$^{-1}$. In this section we generalize
their method in order to account for the bulk velocities and estimate
their impact on the epoch of the first star formation. This problem is
particularly relevant since the effect of the relative velocity on
star formation increases with redshift, and is thus at its maximum
when we consider the very first star. We also study the sensitivity of
the first-star redshift to various uncertainties.

Following \citet{Naoz:2006} we calculate the mean expected number
$\langle N(>z)\rangle$ of star-forming halos that formed at redshift
$z$ or higher, but where the halo abundance is now averaged over the
bulk velocity distribution at each redshift. This number is the
ensemble-averaged number of stars, but we have only one universe to
observe. Hence, we expect Poisson fluctuations in the actual observed
numbers. The probability of finding at least one star is then
$1-\exp\left[-\langle N(>z)\rangle\right]$, and (minus) the redshift
derivative of this gives the probability distribution $p_*(z)$, where
the probability of finding the first star between $z$ and $z+dz$ is
$p_*(z) dz$.

\begin{figure}
\includegraphics[width=3.4in]{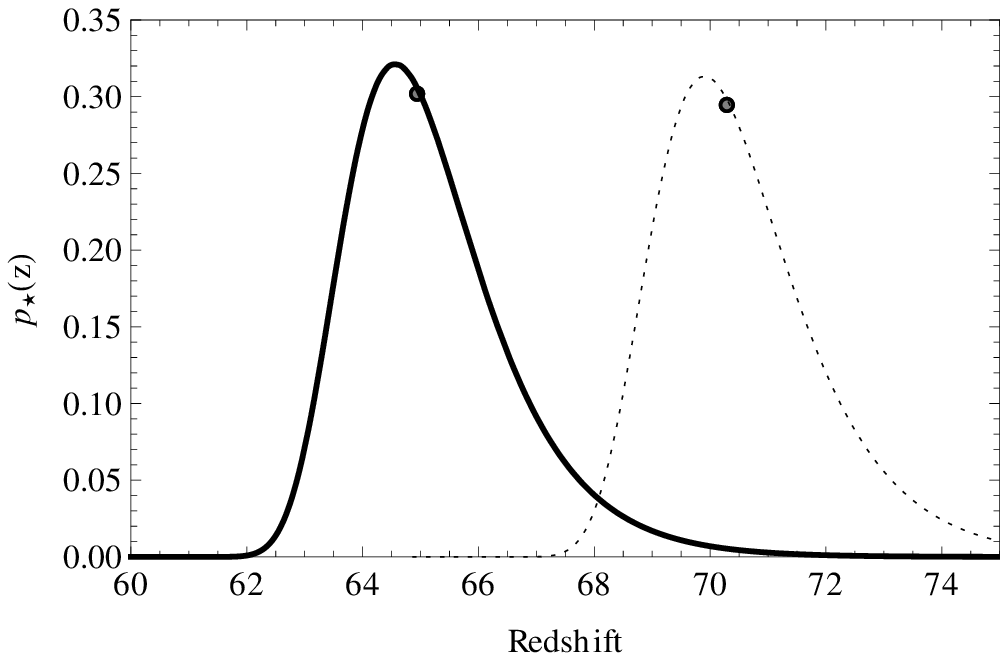}
\includegraphics[width=3.4in]{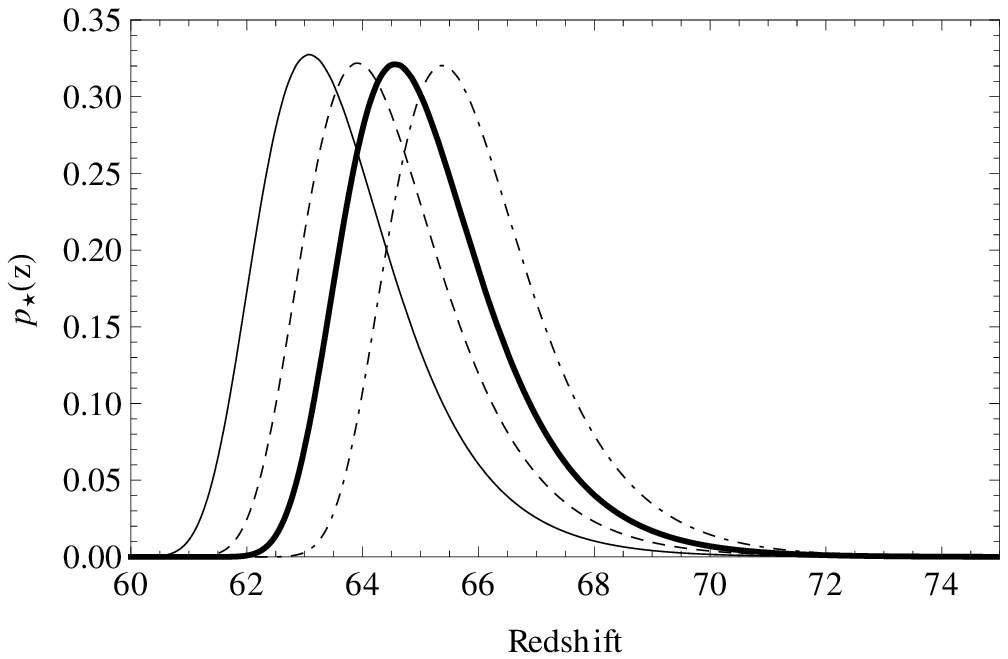}
\caption{\label{Fig:FStar}
{\bf Top panel}: The impact of the relative velocity on the redshift
of the very first observable star. We plot the probability density of
seeing the first star at a given redshift, including the effect of
relative velocity for our optimal fit (solid curve), or without the
effect of the velocity (i.e., for the same fit but with v$_{bc} = 0$,
dotted curve). The formation of the first star is delayed by $\Delta z
= 5.3$ due to the relative velocity effect. We mark the median
redshift of the first star for each distribution ($\bullet$). {\bf
Bottom panel}: The probability density of the redshift of the first
star calculated for each of the fits of Figure~\ref{Fig:Vc}. The
median redshifts of the first star (from left to right) are: $z =
63.5$ (``fit'' to the AMR simulations), $z = 64.3$ (fit to
\citet{Greif:2011}), $z= 65.0$ (the optimal fit to the SPH
simulations) and $z = 65.8$ (fit to \citet{Stacy:2011}).}
\end{figure}

As shown in Figure~\ref{Fig:FStar} (top panel), we find that in the
absence of the bulk velocities, the first star would be most likely to
form at $z=69.9$, with a median $z=70.3$ (corresponding to $t=30$~Myr
after the Big Bang). The difference with \citet{Naoz:2006} is due to
the changes in the cosmological parameters between WMAP3 and WMAP7,
specifically the increased power on the relevant scales (since the
increased spectral index has a larger effect than the reduced
$\sigma_8$), and the decreased cooling mass in the $v\bc=0$ case
compared to the value assumed by \citet{Naoz:2006}.

The relative velocity effect delays star formation, where for the very
first star we find a delay of $\Delta z = 5.3$. The first star is now
most likely to form at $z=64.6$, with a median $z=65.0$ ($t=34$~Myr)
that has a $1-\sigma$ ($68\%$) confidence range $ z = 63.9-66.5$ due
to the Poisson fluctuations. In addition, the redshift of the first
star is uncertain due to the current errors in the cosmological
parameters and the uncertanity in the cooling mass. Regarding the
cosmological parameters, the redshift of the first star is sensitive
to the amount of power on the scale of the first halos. The
uncertainly of WMAP7
\citep{WMAP7} in the amplitude of the primordial fluctuations
(parameterized by $\sigma_8$) is $\Delta
\sigma_8 = \pm0.024$, which implies (for our optimal fit) an
uncertainty of $\Delta z=\pm2.2$ in the median redshift of the first
star. The larger is $\sigma_8$, the earlier will the first star
form. More generally, we include the current correlated errors
in the full suite of standard cosmological parameters, and find
a resulting $\Delta z=\pm5.1$.

In order to estimate the impact of the current uncertainty in the
effect of the bulk velocity on the minimum cooling mass, we estimate
the redshift of the first star for each of the fits discussed in
Section~3. We find (Figure~\ref{Fig:FStar}, bottom panel) that the
range of the SPH simulations is a $\Delta z = 1.5$, and the
discrepancy between the AMR and SPH simulations is comparable. Thus,
we conclude that the delay due to the bulk motion is substantial, but
there are still significant uncertainties in it. In summary, we find
the median redshift of the first star in our observable universe to be
\beq z = 65.0^{+1.5}_{-1.1} {\rm (Poisson)} ^{+0.8}_{-1.5}
{\rm (simulations)} \pm 5.1 {\rm (cosmology)}\ .\eeq Thus, current
uncertainties in the values of the cosmological parameters dominate
over the differences in the simulations and the irreducible Poisson
fluctuations.

\section{Discussion}

We have studied the impact of the relative motion between the gas and
the dark matter on the formation of the first stars. We included a new
effect found in recent small-scale hydrodynamic simulations. In
particular, we fit their results to a physically-motivated ansatz that
expresses the minimum circular velocity of gas-cooling halos as a
simple function of the local bulk velocity when the halo forms. This
result implies that in contrast to previous expectations, the minimum
mass of star-forming halos does not decrease with redshift, except in
regions with very low values of the bulk velocity.

This result implies that the relative velocities produce three
separate effects: suppression of the halo abundance, suppression of
the gas content within each halo, and boosting of the minimum halo
mass required for cooling. Quantitatively, we found that the halo
abundance cut has a large effect on the two categories of halos
(star-forming halos and star-less minihalos), while the cooling mass
boost primarily affects star-forming halos and the suppression of gas
content primarily affects the minihalos. In total, at $z=20$ the bulk
velocities reduce the mean gas fraction in star-forming halos by a
factor of $1.8$ and that in minihalos by $3.1$. Thus, even at $z=20$
the velocity causes order unity fluctuations in the stellar density,
and these fluctuations should be present at the large (100~Mpc) scales
spanned by the velocity correlations.

The velocity dependence of the gas fraction tends to concentrate the
global star formation into regions of low bulk velocity. In
particular, at $z=20$, $68\%$ of the stars are in the $54\%$ of the
volume with the lowest velocity, and $95\%$ are in $89\%$ of the
volume. Adding in the effect of density fluctuations tends to
concentrate the global star formation into regions of both low bulk
velocity and high overdensity. As a result, at $z=20$, $68\%$ of the
stars form within $35\%$ of the volume and $95\%$ in $77\%$ of the
volume. This concentration effect becomes much stronger at higher
redshifts.

The formation of the very first star is delayed by $\Delta z \sim 5$
due to the bulk velocities. Given the updated cosmological and
astrophysical parameters, the first star is now most likely to form at
$z=64.6$, with a median $z=65.0$ (corresponding to a cosmic age of
$t=34$~Myr). Due to the combination of density and velocity
fluctuations, the formation of stars begins at different times in
different regions. This leads to a very inhomogeneous early
universe. Although by redshift 20 most of the universe is populated,
the age of the oldest stars in each region is significantly different.

To make the novelty of our work clear, we now make a full comparison
of the ingredients of our calculations with those in the previous
literature. We start with \citet{Tseliakhovich:2010}, who discovered
that the relative velocity effect is important. They only calculated
the impact on the halo abundance, but this was sufficient for them to
deduce the implication of large-scale fluctuations. However, their
calculations had a number of simplifying assumptions: they calculated
the baryon perturbations under the approximation of a uniform sound
speed, and used the old Press-Schechter halo mass function.

\citet{Dalal:2010} were the first to point out the effect of the
relative velocity on suppressing the gas content of halos. However,
they made a number of simplifying approximations that we have relaxed
here. These include:

(i) We have calculated the filtering mass ($M_F$) from linear theory,
while they took the effective value found in simulations in the
standard (no relative velocity) case, and then multiplied it by a
simple $v\bc$-dependent ansatz.

(ii) We have allowed for a smooth transition between gas-rich halos at
$M\gg M_F$ and gas-poor halos at $M\ll M_F$ as is suggested by
simulations, rather than applying a step-function cutoff.

(iii) We have simultaneously included the dependence of the gas
fraction in halos on the large-scale matter overdensity $\delta_R$ and
relative velocity $v\bc$. This combines both the ``traditional''
biasing model (which includes $\delta_R$ but not $v\bc$) and the
\citet{Dalal:2010} treatment (which includes $v\bc$ but not
$\delta_R$). We found that both effects are important (compare
Sections~4.2 and 4.3).

(iv) We included the effect of $v\bc$ on the halo mass function
\citep{Tseliakhovich:2010}, which \citet{Dalal:2010} did not.

(v) Most importantly, we incorporated a cooling criterion for star
formation, rather than scaling by the total gas content in halos. The
vast majority of the gas is in minihalos that cannot cool, and because
of their low circular velocities their ability to collect baryons is
much more affected by $v\bc$ than the star-forming halos. This
suggests that the effect of relative velocities on early star
formation might be less than found by \citet{Dalal:2010}. However, we
find that the inclusion of the other effects (mass function and
cooling threshold, in addition to baryon fraction) does restore the
expectation for order unity fluctuations, with exciting implications
for observational 21-cm cosmology.

In part of this paper we closely followed
\citet{Tseliakhovich:2010b}. However, we fixed two
inaccuracies in their power spectrum (in the normalization and the spectral
slope) that gave substantially too much power on small scales. Then,
our main goals were to include the new effect on the cooling mass
based on simulations, to extend the calculations to the highest
redshifts of star formation, and to quantify the degree of
concentration of star-forming halos. With there now being three
separate effects of the bulk velocity, we also carefully studied the
relative importance of these various effects.

\section{Acknowledgments}
This work was supported in part by European Research Council grant
203247 (for A.F.) and by Israel Science Foundation grant 823/09 (for
R.B.). D.T.\ and C.H.\ were supported by the National Science
Foundation (AST-0807337) and the U.S.\ Department of Energy
(DE-FG03-92-ER40701). C.H.\ was also supported by the David \& Lucile
Packard Foundation.


\label{lastpage}

\end{document}